# 8 x 8 Terahertz Photoconductive Antenna Array and Parallelized Signal Acquisition System for Fast Spatially Resolved Time Domain Spectroscopy

Henri R.[1], Nallappan K.[1], Graduate Student *Member, IEEE*, Ponomarev D.S.[2,3], Lavrukhin D.V.[2], Yachmenev A.E.[2], Khabibullin R.A.[2] and Skorobogatiy M.[1], Senior *Member, IEEE*

*Abstract*— Terahertz (THz) technology is promising in several applications such as imaging, spectroscopy and communications. Among several methods in the generation and detection of THz waves, a THz time domain system (TDS) that is developed using photoconductive antennas (PCA) as emitter and detector presents several advantages such as simple alignment, low cost, high performance etc. In this work, we report the design, fabrication and characterization of a 2-D PCA array that is capable of detecting both the amplitude and phase of the THz pulse. The PCA array is fabricated using LT-GaAs and has 8 channels with 64 pixels (8x8). The infrared probe beam is steered and focused towards each pixel of the PCA array using a spatial light modulator (SLM). The measured photocurrent (amplitude and phase) from each channel is recorded separately and the frequencies up to 1.4 THz can be detected. Furthermore, the parameters such as directional time delay of the THz pulse, crosstalk between the channels etc., were characterized. Finally, we show that the proposed 2D PCA array design is flexible and can be used for accelerated THz spectral image acquisition.

*Index Terms*— Multipixel detector, Photoconductive antenna array, Terahertz imaging, THz-TDs

## I. INTRODUCTION

TERAHERTZ (THz) frequency band (0.1 to 10 THz) have received a lot of attention in recent years due to increasing number of its technological applications in the fields of telecommunications, imaging and sensing[1-7]. Due to its non-ionizing nature and higher penetration depth when compared to optical infrared (IR) waves, the THz imaging is promising in such applications as security screening[8-10], biomedical imaging[11-16] and industrial process control[17, 18]. Moreover, the THz waves feature higher spatial resolution when compared to microwaves which attracts the development of commercial THz imaging systems[19].

The THz imaging system can be classified into two types: continuous wave (CW) THz imaging system and pulsed THz imaging system (TDS). The CW THz imaging system is a compact, cost-effective, and a relatively high power (compared to THz-TDS) solution for industrial applications in which a real-time THz image at a given frequency can be obtained via raster scanning using a photomixer detector or Schottky detector[20, 21]. Although the CW THz system has higher frequency resolution (~100 MHz), it requires a long acquisition time (several minutes per pixel depending on the integration time and frequency resolution) to obtain a wide frequency spectrum, which limits its application, particularly, in the field of multi/hyperspectral imaging. On the other hand, the THz-TDS is the most commonly used imaging system as it allows to record the broadband THz pulse in a single acquisition. Here, the duration in recording the THz pulse is mainly limited by the speed of the mechanical delay stage. In the THz-TDS, the frequency resolutions are generally in the order of ~1 GHz which also requires smaller step size in the optical delay stage and thereby results in longer acquisition time to obtain an image of the sample/object. Several solutions have been proposed to decrease the acquisition time of the THz-TDS such as asynchronous sampling[22, 23], rotary optical delay line[24], oscillating optical delay line[25], etc. Moreover, even if faster variable optical delay techniques are used, as the power in THz-TDS is spread over a large frequency band, such systems tend to be noisier than the CW THz systems, thus requiring long integration times.

The image acquisition speed in THz-TDS is also limited by the necessity of 2D raster scanning using a single pixel detector such as a PCA [26], non-linear crystals [27, 28] and Air-based-coherent-detectors (ABCD) [29]. By employing mechanical beam steering [30, 31] or compressed sensing techniques such

This work was supported in part by the Canada Research Chair Tier I in Ubiquitous Terahertz Photonics and by the Russian Science Foundation, grant #18-79-10195

[1]*Génie physique, École Polytechnique de Montréal, C. P. 6079, succ. Centre-ville, Montréal, Quebec H3C 3A7, Canada (Corresponding author: maksim.skorobogatiy@polymtl.ca)*

[2]*Institute of Ultra High Frequency Semiconductor Electronics of RAS, 117105 Moscow, Russian Federation (Corresponding author: ponomarev_dmitr@mail.ru)*

[3]*Prokhorov General Physics Institute of the Russian Academy of Sciences, Moscow 119991, Russia*

as phase mask encoding [32], spectral/temporal encoding[33], etc., the image acquisition speed can be improved even when using a single pixel detector.

To further improve the image acquisition speed in THz-TDS, the multipixel detectors such as bolometer arrays[34] and Complementary Metal Oxide Semiconductor (CMOS) cameras[35] are proposed. In these approaches, only the amplitude of the THz electric field is detected, while phase information is lost. In practical applications, however, phase information is highly desirable as it allows direct measurement of material's dielectric constant and/or sample geometry [33].

One of the promising techniques to increase the image acquisition speed and simultaneously to record both amplitude and phase information is to use the PCA array. In 2002, Herrmann et al., proposed and demonstrated the first 1D PCA array with 8 antenna elements separated by 500 µm each [36]. The antenna was fabricated using low-temperature grown GaAs (LT-GaAs) and a particular attention is given to the receiver electronics in which they used alternate pulse locked gated integrator (APOGI) by replacing the commercial multichannel lock-in amplifiers. In 2008, Pradarutti et al., demonstrated the 1D PCA array by increasing the number of antenna elements to 16 [37]. Here, the efficiency of the PCA array was improved by illuminating the probe beam using microlens array and by simultaneously measuring the THz pulse from all the antenna elements. They further increased the number of antenna elements to 128 (1-D) in the following year [38]. It is noted that the incorporation of microlens array is an additional step in the fabrication process which further increases the complexity of the PCA design. In 2014, Brahm et al., demonstrated the fabrication of 1D PCA array (15 antenna elements)[39] using InGaAs/InAlAs multi-layer heterostructure substrate for the excitation wavelength of 1030 nm. All these works show that the fabrication of multipixel PCA array is promising for the development of high-speed THz imaging system. However, the proposed techniques in the literature discussed above lack fast optical beam steering, requires complex wiring to access the individual antenna elements in the PCA array which results in slow mechanical scanning to capture an image.

Another promising technique is the plasmonic nano-antenna array which can be used to increase the sensitivity of the THz pulse detection. In 2016, Yardimci et al., presented a large area detector with a high THz sensitivity (dynamic range of ~90 dB), broadband capacity (~5 THz) and relatively easier alignment[40]. Their 1-pixel design is the main disadvantage for imagery purposes, which limits the speed of the image acquisition. In 2020, Li et al., presented a 63-pixel (9x7) focal plane array of plasmonic nano-antenna arrays [41]. They achieved detection bandwidth of more than 2 THz for all their pixels with the signal-to-noise ratio greater than 60 dB. The main challenge with their design is the complexity of their readout circuit since each pixel is recorded sequentially. Particularly, they used 4 multiplexers and a FPGA development board to route the THz signal from 63 pixels to a single transimpedance amplifier and a commercial single channel lock-in amplifier. Therefore, they could only read 1 pixel at the time resulting in slower THz image acquisition speed.

In this article, we present the design, fabrication and characterization of an 8 channel array of 8x8 PCA antennas (64 pixels) with a goal of its further applications in real-time THz spectral imaging. The 2D PCA array was fabricated using LT-GaAs and characterized for various parameters such as time delay, channel crosstalk etc., using a standard THz-TDS. A spatial light modulator (SLM) is used to focus the IR probe beam on the gap of each PCA antenna elements. This novel approach comes with multiple advantages. One of them is non-mechanic fast optical beam steering with the possibility of using multiple focal spots in order to interrogate multiple antennas simultaneously. In this case, antennas can be grouped in separate channels that can be interrogated in parallel (i.e. by using 'N' transimpedance amplifiers for 'N' number of channels), thus speeding up image acquisition. Moreover, using SLM for interrogation allows smart power budget allocation depending on the total available power of the light source. For example, the number of pixels interrogated in parallel (hence, the number of focal spots generated by SLM) can be increased at the software level when using higher power lasers. At the same time, for a low power laser, one can use just a single beam spot (sequential pixel readout). Furthermore, by combining the proposed technique with the fast optical delay scanning (rotary delay line for example), the image acquisition speed can be further improved.

## II. DESIGN AND FABRICATION OF THZ PHOTOCONDUCTIVE ANTENNA ARRAY

The schematic of the proposed PCA arrays is shown in Fig.

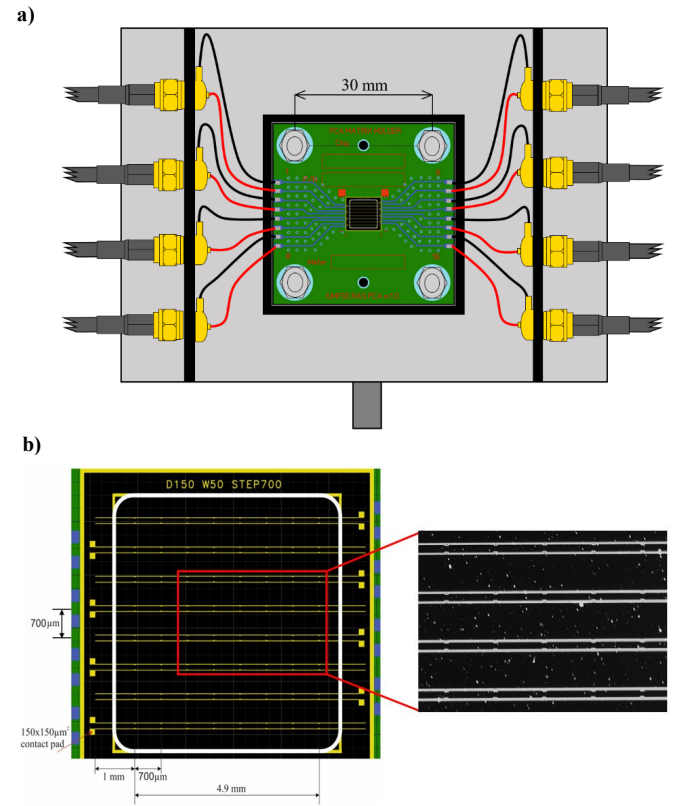

Fig. 1. Schematic of the PCA array for THz detection. **a)** Design of PCA array with a microscopic image of 4x5 PCA of the detector **b)** PCB and holder assembly for the PCA array.

1 (a). The array is comprised of 8x8 dipole H-type PCAs featuring a 100 μm photoconductive gap. The distance between two neighboring bias lines is 150 μm, the area of contact pads is 150x150 μm$^2$. The PCA array was fabricated in two steps. The LT-GaAs was used as a photoconductive substrate. The 0.5 μm thick LT-GaAs was grown via molecular-beam epitaxy using Riber 32P system at 220°C on a semi-insulating GaAs (001) wafer (BEP III/V ratio ~ 35). To improve the LT-GaAs characteristics [42] and isolate the photoconductor from the wafer, a 50 nm thick AlAs layer was sandwiched between the LT-GaAs and the wafer. The LT-GaAs was then annealed at 600°C within 20 min in the growth chamber.

Then, the PCA array was fabricated using an electron-beam lithography (EBL). The sample was dehydrated on the hotplate at 180°C and coated with 540 nm/250 nm thick Copolymer/PMMA 950K bilayer resist stack. The required pattern was exposed via Raith 150-TWO EBL system at 30 kV and developed in an IPA:DI water mixture (1:3) to provide a proper lift-off profile. Then, we used an oxygen plasma descum processing followed by HCl (1:5) solution pretreatment to remove the oxides from the surface of LT-GaAs. Finally, the 50 nm/450 nm thick Ti/Au metallization was deposited on the surface of LT-GaAs via resistive thermal evaporation. The fabrication routine of the single PCA in the array is similar to that reported in our previous paper[43, 44]. The inset in Fig. 1 (a) shows the microscope image of the PCA array. The 6x6 mm$^2$ array was then mounted on a PCB board with a window for THz access on its backside as shown in Fig.1 (b).

On the PCB, there are 16 conductive tracks and each of them is connected to a contact pad. The two conductive tracks associated with one row of PCA are connected to a ring terminal with simple wires. An SMA cable is then used to forward the signal. This means that there are 8 detection channels and that each of them corresponds to a row of 8 PCA.

### III. THz-Time-domain system

We now present the experimental set up that is used for the characterization of the proposed THz PCA array. The schematic of the experimental setup is shown in Fig. 2. A femtosecond laser with the center wavelength of 780 nm, average power of 63 mW, pulse duration of <100 fs and repetition rate of 100 MHz was used as the excitation source. The laser beam was divided into pump and probe beam using a 70:30 beam splitter and a PCA from Menlo system[45] was used as the THz emitter. Finally, a 10-mW pump power was available for the excitation of the THz emitter. As the available pump power was low, only a single spot was used (sequential pixel readout) during PCA interrogation. Two parabolic mirrors were used to collimate and focus the THz beam into the detector with the resultant THz beam spot roughly covering the entire 5mm x 5mm area of the PCA array. When compared with the standard THz-TDS, the pump beam section (towards emitter) is identical while the probe beam section (towards detector) was modified for probing the PCA array. Particularly, a SLM (*Holoeye Pluto with 1920 X 1080 pixels*) [46] was used to steer and to focus the incoming expanded probe beam into the PCA array as shown in Fig. 2. The spot size of the incident probe beam on the SLM was ~8 mm which covers ~1000 x 1000 pixels. The SLM was placed 30 cm away from the PCA array and a single focal spot size of ~200 μm was achieved by displaying a Fresnel lens pattern on the SLM. It is also noted that the higher diffraction of the Fresnel lens pattern was ignored as they carry very low power and their influence in the characterization of PCA array was neglected. The main advantage of using SLM is that a multifocal spot can be trivially created to interrogate all 8 channels simultaneously, and the beam steering can be performed as fast as 60 Hz. Finally, the measured photocurrent from the 8 channels of the PCA array was amplified using 8 dedicated low-noise transimpedance amplifiers from *FEMTO* [47]. A high-speed data acquisition device from *National Instrument* was used to measure the reference signal and the output signals of the 8 transimpedance amplifiers. A multichannel virtual lock-in in *Labview* was developed to extract both phase and amplitude information of the THz pulse. All the wirings are carried out using low noise BNC cables for optimal detection.

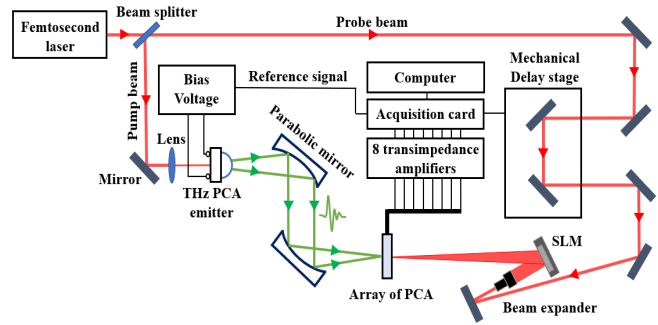

Fig. 2. Schematic of the THz-TDs system for characterizing the PCA Array. A spatial light modulator was used for focusing and steering the femtosecond IR laser beam on each PCA element of the Array.

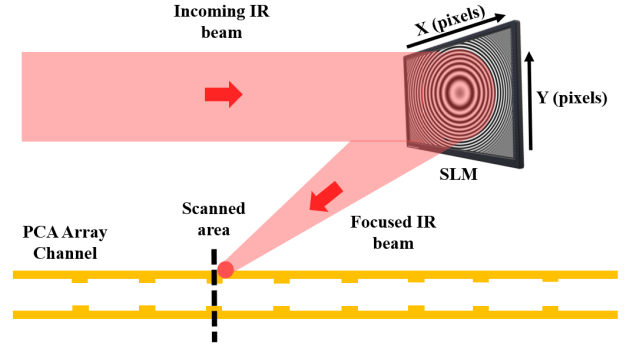

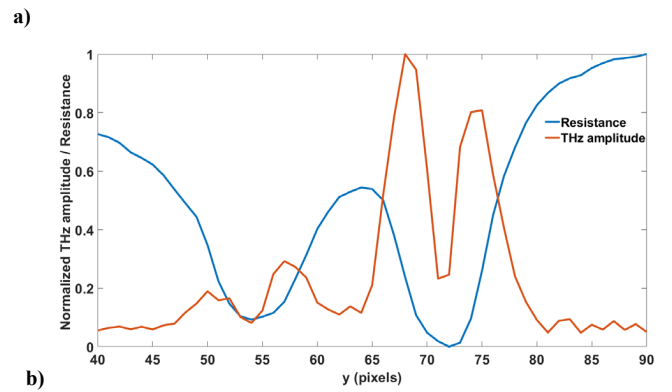

Fig. 3. Calibration and characterization of one of the channels in an 8x8 PCA array **a)** Schematic representation of the experimental setup. The black dotted line represents the scanned area in b). The Fresnel lens pattern is used to focus and steer the IR probe beam. **b)** Resistance and THz amplitude measured while steering the IR probe beam along the Y axis of the detector. Two resistance drops are observed when the IR probe beam is focused directly on the metallic bias lines. Two THz pulses with higher amplitude was observed when the IR probe beam is focused on the edge of the metallic bias line associated with the positive terminal of the transimpedance amplifier and two THz pulses with lower amplitude was observed at the edge of the negative terminal.

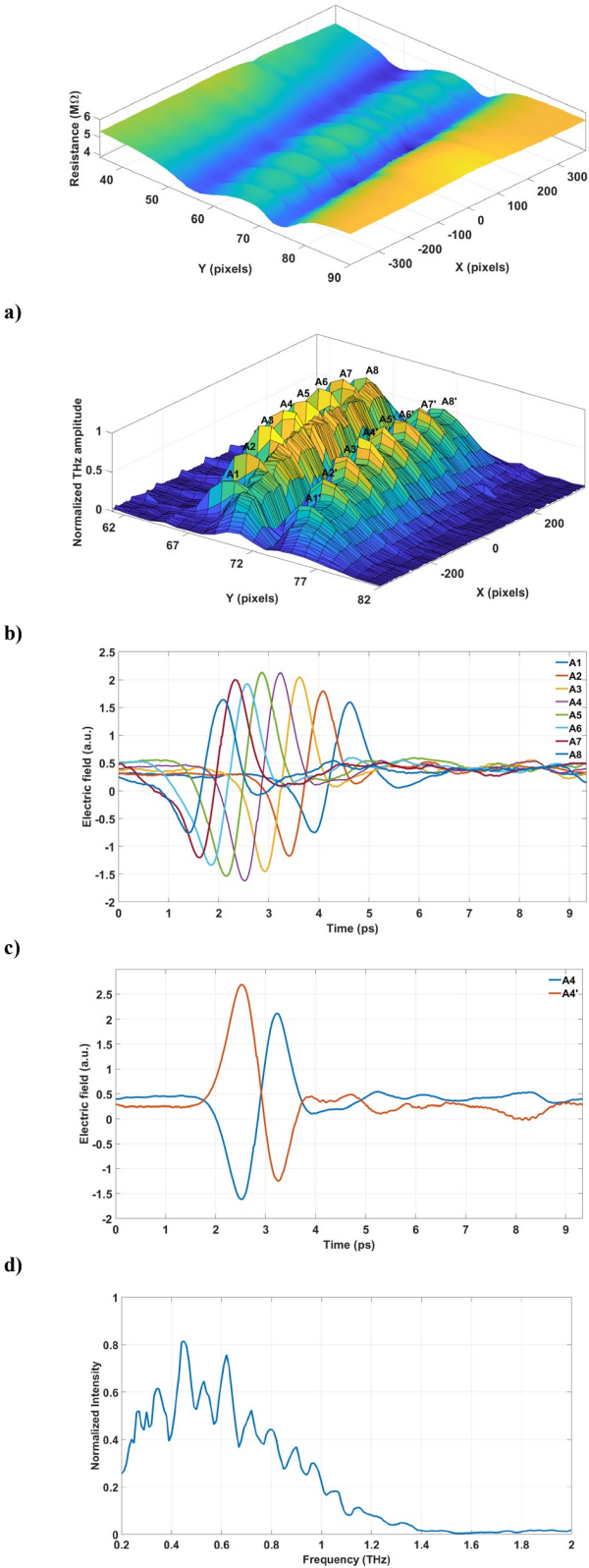

Fig. 4. Calibration and characterization of an 8x8 PCA array. **a)** Resistance map for one of the channels (channel 5) in the array. The X and Y axis represent the position of the Fresnel lens center in the SLM reference frame. **b)** Amplitude of the THz pulse signal for the region associated with the electrode that is connected to the positive terminal of the transimpedance amplifier in a) **c)** THz pulse measured for the position A1 to A8 from b). **d)** THz pulse measured in the position A4 and A4' from b) **e)** Fourier Transform of A4 from b).

## IV. CHARACTERIZATION OF THE THZ PHOTOCONDUCTIVE ANTENNA ARRAY

In this section, we detail the two calibration steps of the proposed PCA array. The goal of the calibration process is to locate the position of the 64 PCA elements in the reference frame of the SLM. By moving the center of the Fresnel lens pattern displayed on the SLM pixel by pixel in both spatial directions (X and Y directions), the focused IR probe beam is steered to scan the whole surface of the PCA array. Therefore, 64 positions of the Fresnel lens pattern on the SLM must be identified, which represent the different positions on the PCA array where maximum THz pulse amplitude is detected. It is noted that the focal length of the Fresnel lens on the PCA array was optimized and remains fixed throughout the calibration process.

The first step in the calibration is to locate each channel of the detector by measuring the drop in the resistivity. Due to the movement of photogenerated charge carriers between the electrodes, the resistance will drop drastically when the IR probe beam is focused near the two electrodes of the channel. Thereby, by steering the IR beam along the Y axis of the detector, one can easily identify the approximate position of each channel. The second step is to locate precisely where the THz pulse is detected with a maximum amplitude. By moving the IR probe beam across the PCA array and recording the THz signal for each position, it is possible to identify the exact location where the detection of the signal is optimal. As recording the THz pulse takes longer duration due to the slow scanning speed of mechanical delay line, measuring the resistance first is a rapid method to identify the region of interest on the PCA array.

The schematic of the calibration process is shown in Fig.3 (a), where the incident IR probe beam on the SLM was focused on to the PCA array. One pixel in the SLM reference frame corresponds to 7μm in the PCA array. This is identified based on the known dimensions of the PCA array (100 μm gap at the H-structure and 150 μm spacing between the electrodes). In Fig. 3 (b), we present the normalized value of the drop-in resistance and the THz pulse amplitude measured across the black dotted region (Y-direction of channel 5) described in Fig. 3 (a). We observe two positions where the resistance drops and four peaks for the THz pulse amplitude. The resistance drops when the IR beam was focused on the electrodes. The THz pulses were recorded when the IR beam was focused at the edges of the electrodes (on both sides) and hence we measure four maxima in THz amplitude. The amplitude of the THz pulse depends on the polarity of the electrode that is connected to the transimpedance amplifier. This was verified by reversing the polarity of the electrode that is connected to the transimpedance amplifier. This was verified by reversing the polarity of the channel and observing the two highest THz amplitude peaks switching to the adjacent 2-peak group. It was reported that, in addition to the structure of the antenna, the lifetime and mobility of the charge carriers play a significant role in the detection of THz signal [48]. In LT-GaAs, the electron and hole carrier lifetimes are 0.1 ps and 0.4 ps, and the initial electron and hole mobilities are 200 cm$^2$/V s and 30 cm$^2$/V s respectively. Due to the difference in the lifetimes and mobilities of electrons and holes, the detection at the electrode that is connected to the positive terminal of the transimpedance amplifier will be predominant. This phenomenon is known for PCA emitter where the THz power is maximal when the IR beam is focused close to the anode[49-53].

The final step in the calibration process is to precisely locate the 8 antenna elements of each channel (2D). As an example, the resistance map measured in the channel 5 of the PCA array as the region of interest is shown in Fig. 4 (a). Similarly, the two THz peaks associated with the electrode that is connected to the positive terminal of the transimpedance amplifier were measured and the corresponding amplitude map is shown in Fig. 4 (b). It should be noted that the amplitude of the THz pulse is smaller at the edges of the channel (X-direction) when compared to its center because of the low power of the IR probe beam at the edges due to a higher diffraction order of the Fresnel lens pattern and also due to the gaussian distribution of the THz beam. We can also observe that THz signal can still be detected in between the 8 H-gap antennas (A1 to A8). Although the signal is weaker, it could be possible to increase the resolution of the detector by measuring the signal where there is only the two gold bias lines separated by 150 μm. From the two rows of THz pulse amplitudes marked in Fig. 4 (b), the time-resolved THz pulses corresponding to one of the rows is shown in Fig. 4(c). We observe a relative time delay in the THz pulses that is recorded along the different positions of the channel. The ability of a 2D antenna array to measure spatial distribution of a pulse delay can be used for the THz radar applications to locate the direction and extrapolate the location of a THz light source.

In Fig. 4 (d), we present the time-resolved THz pulses for the position A4 and A4' (4$^{th}$ antenna element of channel 5) as marked in Fig. 4 (b) which correspond to the IR laser beam being focused on both the edges of the electrode that is connected to the positive terminal of the transimpedance amplifier. In Fig. 4(e), we present the Fourier transform of one of the THz pulses measured with the array detector. The THz frequencies up to 1.4 THz can be detected. The oscillations in the spectrum can be explained by the Fabry-Perot effect due to resonances of the THz light inside the detector substrate. In classical one-pixel PCA detector, generally a silicon lens is placed on the detector to focus the THz light into the antenna substrate, which also helps with decreasing the effect of back reflections. To reduce back reflections in the antenna array, an 8x8 micro-lens array could be designed to focus the THz light on each PCA of the detector, which is reserved for our future work.

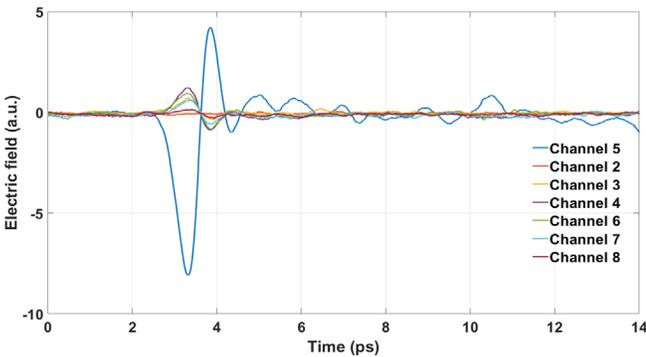

Fig. 5. Measurement of crosstalk for the detector array. The IR beam is focused on to a single antenna element (#4) of the channel 5. The signal measured at the neighboring channels (2, 3, 4, 6, 7 and 8) are shown.

In array-based detectors, one of the major concerns is the channel crosstalk between the neighboring channels which limits the sensitivity of the system. Therefore, in the following, we measure the crosstalk between the channels of the proposed PCA array. In order to estimate the channel crosstalk, the IR probe beam was focused into the fourth dipole antenna of the fifth channel located near the center of the PCA array. Then, the THz signal was measured at the neighboring channels as shown in Fig.5. From the figure we see that the maximal crosstalk of 15% is observed for the immediate neighboring channels (channel 4 and channel 6), while the crosstalk decreases rapidly for the channels that are farther away. The channel crosstalk can be further minimized computationally after measuring all 64 spectrally resolved crosstalk matrices by exciting one antenna at a time and measuring crosstalk at the location of other antennas.

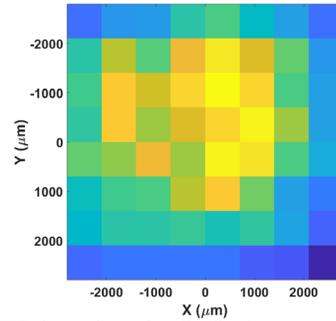

Fig. 6. Raw THz beam intensity distribution measured with the 8x8 PCA array.

In order to use the 8 channels of the detector for imaging, an array calibration is necessary. First, one must find the position of the 64 pixels (defined as positions of the local maxima in the THz amplitude), and then record the reference THz pulses under illumination conditions that will be later used in imaging. Final amplitude, phase and spectral images can be retrieved by Fourier transforming the detected THz pulses after insertion of an object under imaging, and then dividing such spectral intensities by the corresponding ones of the reference. In Fig. 6, an example of such reference is presented. The diameter of the THz Gaussian beam is 3 mm at the focal point which corresponds to the maximal amplitude of the THz wave in the frequency domain. Although, the THz image obtained is pixelated, the resolution could be improved by increasing the number of pixels in the channel since THz pulse can be detected in between the antenna as mentioned before.

## V. CONCLUSION

To conclude, in this work, we presented the design, fabrication and characterization of a 2D 8x8 Photoconductive Antenna array for THz detection. With such arrays both amplitude and phase of the THz signal can be recorded. High-speed THz image acquisition can be achieved due to multi-channel parallel data acquisition using virtual lock-in, as well as fast multi-beam steering capability of the SLM (60 Hz). The imaging speed can be further improved by replacing a slow linear optical delay line with the fast rotary delay line. Therefore, we believe that the proposed PCA array can be of practical importance in the development of next generation of fast spectral imaging systems in the THz range.

ACKNOWLEDGMENT

This work is funded in part by the Canada Research Chair I in Ubiquitous THz photonics of Prof. Maksim Skorobogatiy. The development and fabrication of the PCA array was carried out within the support of the Russian Science Foundation, grant #18-79-10195.